\documentclass[aps,pre,twocolumn,superscriptaddress,nofootinbib,longbibliography]{revtex4-1}

\usepackage{amsthm}

\usepackage[T1]{fontenc}
\usepackage[utf8]{inputenc}

\usepackage{graphicx}
\usepackage{wrapfig}

\usepackage{framed} % or mdframed or tcolorbox for more control
\usepackage{hyperref}

\usepackage{xcolor}
\definecolor{MyGray}{rgb}{.93,0.93,0.93}
\definecolor{RoyalBlue}{RGB}{65,105,225}

\usepackage{amsmath,amssymb}
\usepackage{calrsfs}

\usepackage{comment}
\usepackage{ifthen}
\usepackage{todonotes}



\begin{document}

% Title
\title{Bifurcations and Phase Transitions in the Origins of Life}

% Author definitions
\author{Ricard Solé}\email{ricard.sole@upf.edu}
\affiliation{Complex Systems Lab, Universitat Pompeu Fabra, 08003 Barcelona, Spain}
\affiliation{Institut de Biologia Evolutiva, CSIC-UPF, 08003 Barcelona, Spain}
\affiliation{Santa Fe Institute, 1399 Hyde Park Road, Santa Fe, NM 87501, USA}

\author{Manlio De Domenico}\email{manlio.dedomenico@unipd.it}

\affiliation{Complex Multilayer Networks Lab, Department of Physics and Astronomy 'Galileo Galilei', University of Padua, Via Marzolo 8, 35131 Padova, Italy}
\affiliation{Istituto Nazionale di Fisica Nucleare, Sez. Padova, 35131 Padova, Italy}
\affiliation{Padua Center for Network Medicine, Via Marzolo 8, 35131 Padova, Italy}
\affiliation{Padua Neuroscience Center, Via Giuseppe Orus, 2, Italy}

\date{\today}

\begin{abstract}
The path toward the emergence of life in our biosphere involved several key events allowing for the persistence, reproduction and evolution of molecular systems. All these processes took place in a given environmental context and required both molecular diversity and the right non-equilibrium conditions to sustain and favour complex self-sustaining molecular networks capable of evolving by natural selection. Life is a process that departs from non-life in several ways and cannot be reduced to standard chemical reactions. Moreover, achieving higher levels of complexity required the emergence of novelties. How did that happen? Here, we review different case studies associated with the early origins of life in terms of phase transitions and bifurcations, using symmetry breaking and percolation as two central components. We discuss simple models that allow for understanding key steps regarding life origins, such as molecular chirality, the transition to the first replicators and cooperators, the problem of error thresholds and information loss, and the potential for "order for free" as the basis for the emergence of life. 
\end{abstract}

\maketitle

\tableofcontents

%%%%%%%%%%%%%%%%%%%%
%%%%%%%%%%%%%%%%%%%%
%%%%%%%%%%%%%%%%%%%%

\section{Introduction}

The transition from non-living to living matter remains a scientific puzzle. How can we move from the first to the second, where life can be understood as a dynamic process associated with discrete, self-replicating systems? Is the origin of life one particular case study within evolutionary theory? Charles Darwin \cite{pereto2009charles} referred to the problem within {\it The origin of Species} in these terms in a letter to Joseph Hooker (1871): 
\begin{quote}
    \it "But if (and oh what a big if) we could conceive in some warm little pond with all sorts of ammonia and phosphoric salts, light, heat, electricity, etcetera present, that a protein compound was chemically formed, ready to undergo still more complex changes [..] "
\end{quote}
As described, Darwin understood that a crucial aspect of our quest for how life emerged and evolved would require bringing together chemistry and life. Nowadays, the origins of life remain unsolved and are often controversial \cite{pereto2005controversias,chyba2005astrobiology}. In particular, the rarity or inevitability of life has been under dispute. Alexander Oparin, one of the founder fathers of the origin of life studies, said \cite{oparin1975theoretical}: 
\begin{quote}
    \it "
There is every reason now to see in the origin of life not a "happy accident" but a completely regular phenomenon, an inherent component of the evolutionary development of our planet. The search for life beyond Earth is thus only a part of the more general question which confronts science, the origin of life in the universe.
 "
\end{quote}
Others have argued that highly contingent conditions might have been required to evolve early life\cite{monod1974chance,bertrand2023controversy}. One particular approach is based on Drake's equation \cite{vsklovskij1966intelligent,cirkovic2004temporal}
to find life under a set of $n$ constraints weighted by their probabilities $p_{\mu}$. This formula provides a probability that intends to incorporate each level within a multiplicative function $P_{\text{life}}=\prod_{\mu=1} p_{\mu}$ where each extra term reduces the probability by an order of magnitude. It could be argued that adding extra terms regarding required molecular innovations (such as the formation of long polymers or stable metabolisms) might strongly reduce the chances of life from happening.

How can we make inferences about the origins of life, given the vast chasm of time since it took place and the possible contingent nature of the underlying processes? These questions are tied to the problem of the historical nature of molecular evolution: are there multiple paths to the emergence of life on our planet and elsewhere? \cite{grefenstette2024chapter}. Or, as suggested by other studies, are there universal properties of complex systems that constrain the space of the possible? \cite{sole2024fundamental}. Since we have only one large-scale natural experiment that leads to life in our biosphere, it might seem impossible to answer these questions from an inferential perspective due to the lack of data. 

Life elsewhere could be so different that its underlying principles have nothing to do with our understanding of living matter. This is the problem of "life as we don't know it" \cite{wachtershauser2000life,cleland2019quest,grefenstette2024chapter}: can we articulate a universal approach that could safely explain the properties of unknown life? These are important questions, particularly within the growing field of astrobiology \cite{schopf1999cradle,marais1999astrobiology,chyba2005astrobiology,smith2021grayness,asche2023takes,kempes2021multiple}. The potential diversity of life forms that could have emerged and evolved elsewhere raises two relevant questions. The first is how we can study potential scenarios for the origins of life using experiments. The second question is what kind of theoretical approach can capture the universal properties of the problem. Let us consider them.

Concerning the first problem, although alternative life forms are likely to exist in exoplanets (or even in our solar system), detection is a big challenge and a central problem for astrobiology \cite{lovelock1965physical,lovelock1974atmospheric,sole2004large,schwieterman2018exoplanet,walker2018exoplanet,marshall2021identifying}. However, we have an alternative way to look at the problem of the origins and evolution of life based on a "synthetic" approach \cite{sole2016synthetic,luisi2016emergence,porcar2014synthetic}. In summary, we could explore the problem of past events concerning the emergence of living complexity using experimental implementations of prebiotic scenarios or through {\it in silico} models of artificial life. The first path has a long tradition and has shaped the early developments in the field by its pioneers \cite{tirard2010origin,schopf2024pioneers}. In particular, the classical Stanley Miller experiment that revealed the prebiotic synthesis of aminoacids \cite{miller1953production,bada2003prebiotic} and Joan Or\'o's synthesis of adenine \cite{oro1960synthesis,oro1961mechanism,basile1984prebiotic}.On the other hand, artificial chemistries can also be defined to explore real and alternative possibilities \cite{banzhaf2015artificial,rasmussen2019toward}. These experiments have been instrumental in developing our understanding of the generative potential of these surrogates of alien prebiotic worlds.

What about the "right" theoretical framework? The study of the origins of life has greatly benefited from the development of theoretical models, some of which we discuss below. One particularly important contribution concerns our understanding of the mathematical principles associated with the dynamics of replicators and reproducers in terms of population dynamics \cite{szathmary1997replicators}, while other authors have explored the problem within the context of information \cite{eigen1981origin,kuppers2012molecular,walker2013algorithmic,walker2017matter}. Ideally, any satisfactory approach should be able to deal with the potentially enormous richness of physical, chemical, and geochemical complexity and diversity of possible worlds. As with other aspects of evolutionary biology, the correct view seems to fall at the crossroads between history and physics \cite{smith1982between}. Importantly, the emergence of living matter required several key transitions that allowed in particular: (a) the creation of useful chemical diversity, (b) the shift from competitive to cooperative phenomena that allowed molecular complexity to grow, and (c) the emergence of life precursors associated with reaction and metabolic networks. These are the three case studies considered here. It is remarkable that, in all these cases, we can define dynamical models whose analysis reveal well-defined explanations for non-trivial properties of reality. 

Two related fields provide the best candidates: theoretical models based on phase transitions and nonlinear dynamical systems. In both cases, emergent patterns result from the non-linearity of interactions among different system components. This concept has been applied to the emergence of new classes of biological complexity in evolution (the so-called major evolutionary transitions) \cite{szathmary1995major,schuster1996does,lane2010life,de2002life} and an explicit connection to phase transitions and collective phenomena elaborated by several authors \cite{kauffman1995home,szathmary2015toward,sole2016synthetic,wolf2018physical,goldenfeld2011life}. The conceptual framework of phase transitions (which we summarize in the following) has been a successful approach to understanding reality across scales, showing how new phenomena are captured by very simple models capable of explaining how complexity emerges. Three key ingredients, namely symmetry breaking, bifurcations, and percolation, will play a leading role. We aim to connect formal concepts from physics and nonlinear dynamics—typically presented in highly technical textbook discussions—with the evolutionary transitions explored in our case studies. To achieve this, we focus on the simplest models that effectively capture the key phenomenological aspects of each problem, often setting aside potentially relevant molecular-level features.

%refs: Koonin Big Bang, refs on major transitions (Lane, de Duve..)

\section{Modelling phase transitions}

Along with experimental approaches based on artificial chemistries emerging from potential prebiotic conditions, the field of life origins has benefited from a rich tradition of theoretical models. The models range from more or less standard population dynamics equations (allowing competition, replication, and mutation) to sophisticated models where microscopic rules and environmental factors have been incorporated. Despite their diversity, most of them fall into two major classes of models: statistical physics approaches or nonlinear dynamical models.
In this section, we provide a brief theoretical and computational analysis of phases and phase transitions using simple but powerful models that will help to develop our examples in the following and illustrate the similarities and differences between the two descriptions.

\subsection{Physics of phase transitions}

One familiar example of phase transitions is given by boiling water, as shown in Fig. 1a, when $T \sim 100^o C$. Two phases are visible: water (the liquid phase) and steam bubbles (the gas phase). Before achieving this temperature, the water remains liquid, as a single phase, within a very wide temperature range, for $0<T<100$. A phase transition occurs at $T_c^{boil}$, when water transforms into gas, showing a $1800$-fold decrease in density (where steam bubbles coexist with the liquid). The dramatic change requires a slight increase in $T$, named the {\it control parameter}. On the other hand, density $\rho$ quantifies the system's response and is the so-called {\it order parameter}. 

The response of magnets (Fig 1b) provides one seemingly different example of a transition to changes in temperature. In a nutshell, if we measure the strength of the magnet or {\it magnetization} $M$ and plot it against the external temperature (Fig.2c), we can see that a sharp change occurs close to a given critical value $T_c$, known as the Curie temperature. Below $T_c$, the material is ferromagnetic: its units or {\it spins} (tiny magnets) align in the same direction, creating a strong overall magnetization. Magnetization is just how strongly a material acts like a magnet. Above the Curie temperature, the material becomes paramagnetic, with spins adopting random directions because of heat, and the overall magnetization disappears. The filled circles in Fig.1c are experimental results of a 2D magnetic material \cite{back1995experimental}. A model of such a system is the Ising model, sketched in Fig.1d as a square lattice, where each node is a spin and interacts with its four nearest neighbors. 

The Ising model allows for a statistical approach to complexity, phase transitions and emergent phenomena \cite{yeomans1992statistical,wilson1979problems,artime2022origin}. It has been validated across scales and systems, from cell biology \cite{duke1999heightened}, multicellularity \cite{weber2016cellular} and ecology \cite{schlicht2004forest} to the economy \cite{ball2006critical,sornette2014physics}. Briefly speaking, statistical physics aims to understand the macroscopic properties of a system (such as density or magnetization) by studying the collective behaviour of its microscopic components (such as atoms, molecules, or cells). It uses probabilistic methods to describe how these components interact and change, connecting microscopic dynamics to observable phenomena through averages and distributions. One of the greatest achievements of this framework is the realization that the dynamics of the system is such that very simple models, devoid of most details associated with the underlying units, can fully account for the different phases and the associated transitions. 

  %@book{ball2006critical, title={Critical mass: How one thing leads to another}, author={Ball, Philip},  year={2006}, publisher={Farrar, Straus and Giroux}}

\begin{figure*}[t]
{\centering 
\includegraphics[width=\textwidth]{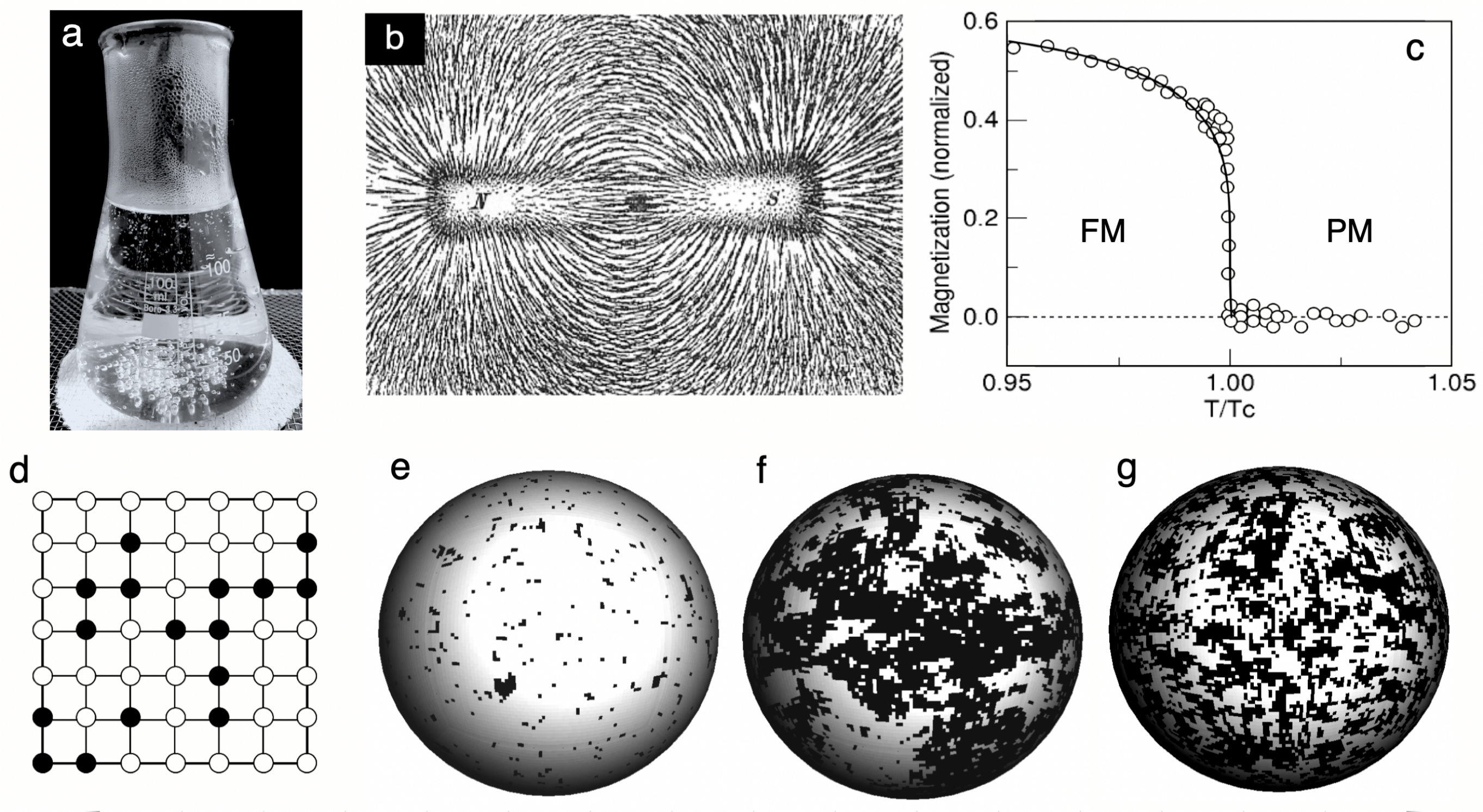}
\caption{Phase transitions in physics. Two examples are: (a) the water-steam (liquid-gas) transition that occurs when water is heated beyond $T_c=100º$ and (b-c) the ferromagnetic transition, where a magnet (b) loses its magnetization properties as shown on (c) where $M$ is measured (open circles) as $T$ grows beyond a critical value $T_c$ for a two-dimensional (atomic) layer of iron (adapted from \cite{back1995experimental}). The two phases are ferromagnetic (FM) and paramagnetic (PM). A toy model used to explain this transition is provided by the two-dimensional Ising model, represented in (d) as a lattice where atoms (spins) have two possible states: "up" ($\uparrow$) and "down" ($\downarrow$), indicated as black and white circles, respectively. Spins only interact with their nearest neighbours. Three examples of the model dynamics at three different temperatures are shown for (e) a sub-critical ($T<T_c$), (f) critical ($T=T_c$ and supercritical ($T>T_c$) are displayed.}
}
\end{figure*}  

The basic setting of the model is a spin lattice, where each spin $S_i$ can take a value of $+1$ or $-1$. A {\it microstate} of the system, i.e. a specific configuration of $n$ spins, is denoted as $\boldsymbol{\sigma}=\{S_{1},S_{2},...S_{n}\}$. As standard in statistical mechanics, the energy of the system is given by the Hamiltonian (an energy function)
\begin{equation}
  {\cal H} (\boldsymbol{\sigma}) = -{1 \over 2}\sum_{\langle i,k \rangle} J S_k S_i.  
\end{equation}
The coupling constant $J$ defines the strength of the interaction between nearest neighbours $\langle i,k \rangle$ within the lattice of magnets\footnote{A more complete form of its energy function involves two terms, namely: 
$\mathcal{H}(\boldsymbol{\sigma}) = -J \sum_{\langle i,j \rangle} S_i S_j/2 - h \sum_{i} S_i$
where $h$ is an external magnetic field.}. It can be shown that, for a given temperature $T$, the system is found in an equilibrium configuration $\boldsymbol{\sigma}$ given by the Boltzmann probability 
$$P(\boldsymbol{\sigma})={1 \over Z}e^{-\beta\mathcal{H}(\boldsymbol{\sigma})}$$
where $\beta=1/k_{B} T$ is the inverse thermal energy, $k_{B}$ is the Boltzmann constant and $Z=\sum_{\boldsymbol{\sigma}}e^{-\beta\mathcal{H}(\boldsymbol{\sigma})}$ is the so-called partition function. As we can see, configurations with lower energy have higher probabilities, as $\exp(-\beta \cal H)$ is larger for smaller energies. For low temperatures ($T \rightarrow 0$, or $\beta \rightarrow \infty$) the system will approach its ground state (minimal energy configuration) whereas when temperatures are high, all configurations will be (nearly) equally probable. These two limits, as will be seen below, correspond to ordered and disordered phases. How can we model these phases and their time dynamics?

The kinetic Ising model is a dynamical extension incorporating time evolution and non-equilibrium behavior. It describes how the system evolves under specific microscopic dynamical rules. Dynamical changes in the orientation of the magnets are introduced using transition probabilities that depend on temperature and the neighbor spins. The goal is to provide a rule of change associated with single spins, and one way of doing it is to compute the  transition probability: 
\begin{equation}
W[S_i \rightarrow -S_i] = {1 \over 1 + e^{\Delta  {\cal H}/k_B T}}, 
\end{equation}
where $\Delta{\cal H}$ is the energy change associated with a spin flip, which is given by $\Delta  {\cal H}=2J\sum_{\rangle i,j \langle}S_iS_j$. By using this expression, we can update the state of individual spins and study the relaxation dynamics of the whole lattice, as given by the magnetization $M = \frac{1}{n} \sum_{i=1}^{n} \langle S_i \rangle$. Despite being a toy model, the model not only predicts the presence of a phase transition; the predicted magnetization matches very well the observed experiments (Fig. 1c, continuous line). It also predicts with enormous precision the statistical properties of the real material at $T=T_c$. Three snapshots of this model are shown in Fig.1e-g for subcritical, critical, and supercritical temperatures. An important observation needs to be made here. At the subcritical phase, where spins are aligned, two possible macroscopic states are equally possible ("all up" or "all down"). In Fig. 1e we only display one of these two possible symmetric states (they have the same energy). Once we cross the critical temperature from higher to lower values, the chosen macroscopic state is the result of chance. This behavior is a hallmark of {\it spontaneous symmetry breaking}, where the system selects a particular ordered state from multiple equivalent possibilities, leading to collective behaviour emerging from local interactions. Moreover, very complex structures can be observed at $T_c$, where we can appreciate the emergence of self-similar patterns in space and time \cite{wilson1979problems}. 

Studying these critical points is a fundamental part of statistical physics and requires some advanced techniques. However, we can study phase transitions using a different approach, which we will exploit here. This will allow us to derive several phenomena that are deeply connected with key steps towards life origins using simple mathematical models.

%======================================================

\subsection{Bifurcations and mean field models}

%======================================================

In statistical physics, macroscopic properties (such as magnetization) emerge from the collective behaviour of microscopic states. Macroscopic states are described using order parameters that summarize the system's state. To obtain simpler models, we can replace detailed interactions with an average field experienced by each component. This reduces the many-body problem to an effective single-body problem governed by self-consistent equations. The dynamics of these equations can often be defined in terms of low-dimensional dynamical systems. This approach bridges microscopic details with macroscopic behaviour through a simplified yet insightful framework.

Let us illustrate this by using again the Ising model. The idea of mean-field approximations is that we can ignore spatial correlations by assuming that the behaviour of a given unit is a function of the average system's state (and not the state of its nearest neighbours). We want to know the expected equilibrium states, defined by our order parameter, as a control parameter is tuned.  
By ignoring the local correlations, we end up with a simplified picture that captures the key qualitative properties of the phase transition. For a broad class of models, including Ising, we can describe these properties using an effective framework as a continuous approximation:
\begin{equation}
{dM \over dt}=F(M;T)=(T_c-T) M - M^3
\end{equation}
i.e. a nonlinear ordinary differential equation (ode). We will use this example as a case study to illustrate how this class of mathematical models is studied using linear stability analysis~\cite{strogatz2018nonlinear}. The first step is to determine the set of fixed points, i.e., those $M^*$ values such that the equilibrium condition
\begin{equation}
\left ( {dM \over dt} \right )_{M^*}=0
\end{equation}
holds. In simple terms, this condition tells us that the state of this system does not change with time. It can be easily shown that there are three fixed points, i.e. $M^* \in \{ 0, T_{\pm}=\pm \sqrt{T_c-T} \}$. How do they behave? In general, different types of fixed points are possible. Intuition is captured in a physical analogy shown in Fig. 2a, where we display three marbles sitting on different landscape locations. All of them are in equilibrium, but the nature of these equilibria is different. Those at the bottom of the valleys are {\it stable}: a small perturbation in their position makes them roll back to the bottom, where they will stay. By contrast, the one sitting at the top of the hill will leave it, even if a slight perturbation is applied: they are {\it unstable}\footnote{In physics, this stability condition emerges from an energy landscape picture, where the surface defining our landscape has an associated potential energy function $U$. In one dimension, where $U=U(x)$, the force $F(x)$ associated with the landscape where a particle sits will be opposite to the slope of $U$, that is, $F=-dU(x)/dx$.}. It can be shown that the sign of the Lyapunov parameter $\lambda$, defined as:
\begin{equation}
\lambda(M^*) = \left (  {dF(M;T) \over dM} \right )_{M^*} = (T_c-T+M^3)_{M^*}
\end{equation}
 provides the condition for stability: if $\lambda(M^*)<0$, the point will be stable, whereas for $\lambda(M^*)>0$, $M^*$ will be unstable. For $T>T_c$, the only stable state is $M^*=0$, while for $T<T_c$, the zero state is unstable, whereas the two symmetric solutions $T_{\pm}$ exist and $M^*=0$ is now unstable. In our context, the transition is known as a {\it bifurcation}. Generally, a bifurcation refers to a qualitative change in the system's behaviour or structure as a parameter crosses a given (critical) value. This can involve the creation, destruction, or change in stability of fixed points, periodic orbits, or other attractors. This bifurcation leads to a broken symmetry for our specific case, as summarized in Fig. 2c-d. The global behaviour of our dynamical system is fully captured by the so-called {\it bifurcation diagram}, where the different fixed points (attractors) and their stability are displayed against the external parameter. In this diagram, stable and unstable attractor sets are indicated by continuous and discontinuous lines, and this particular bifurcation leading to two stable states is known as a {\it pitchfork bifurcation} \cite{sole2011phase}. This kind of bifurcation maps, within dynamical systems, is the symmetry-breaking scenario displayed by the Ising model.

\begin{figure*}[t]
{\centering 
\includegraphics[width=\textwidth]{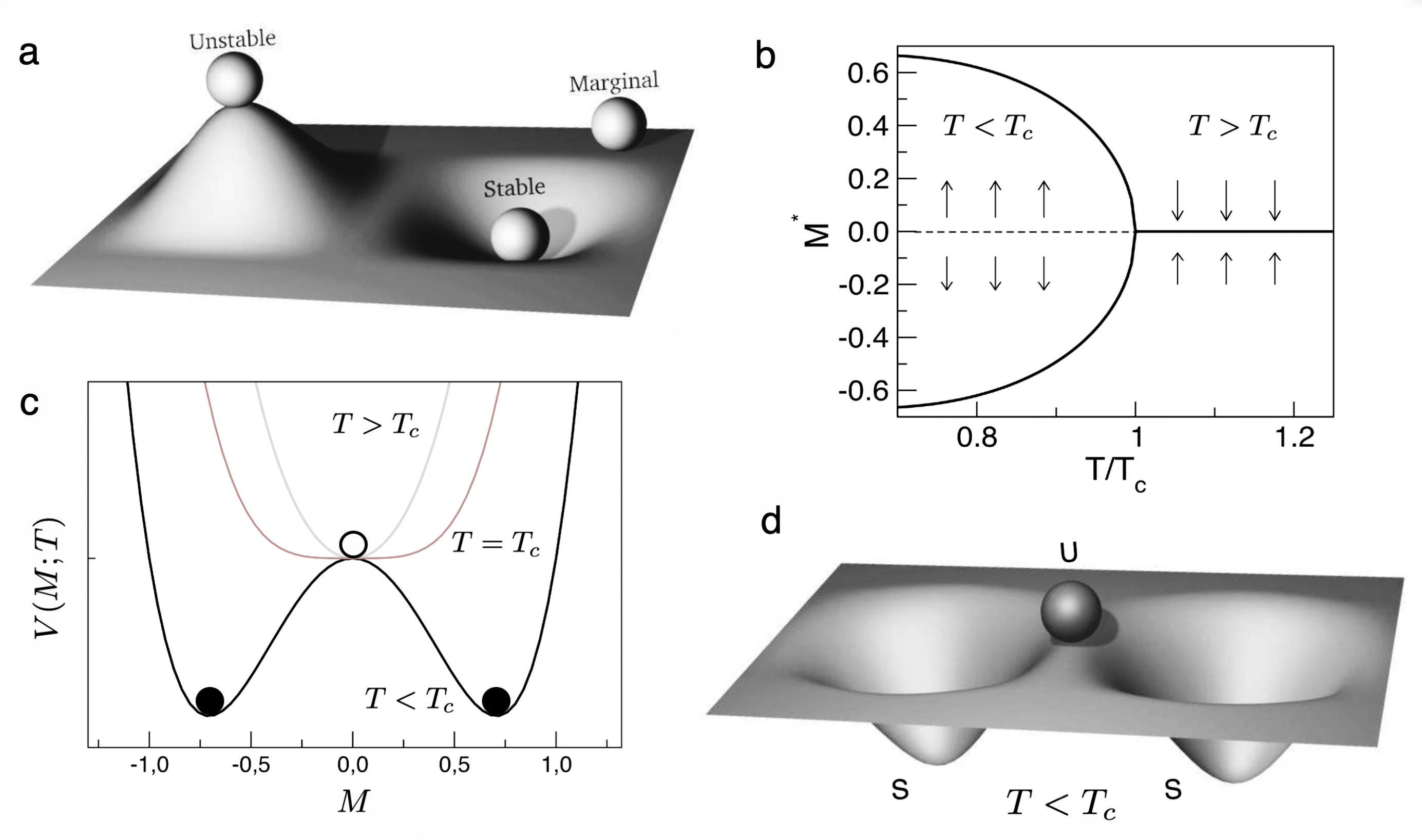}
\caption{Phase transitions as bifurcations in dynamical systems. In (a), a physical analogy is used to illustrate the concept of different kinds of equilibria. For the mean field approximation of the Ising model, the bifurcation diagram for $M$ against the normalized temperature (b) reveals a pitchfork bifurcation with two symmetric branches below the critical temperature. The associated potential function (a surrogate of the energy function of the Ising model) is shown in (c) for different values of $T$. The mechanical analogy with this problem is displayed in (d). }
}
\end{figure*}

Can we formalize the previous landscape of marbles rolling on a surface for our dynamical system? 
The answer is yes for one-dimensional dynamical systems like the one described here. A potential function, see \cite{strogatz2018nonlinear}, can be derived from 
\begin{equation}
{dM \over dt}=-{dV(M)\over dM}
\end{equation}
where we assume that the dynamics (i.e., the change over time of $M$ given by $\dot{M}$) can be derived from the gradient of a potential $V(M)$. From the definition, it is easy to see that
\begin{equation}
V(M)=-\int \dot{M}dM
\end{equation}
which, in this case, gives a fourth-order expression that is a symmetric function, i.e.:
\begin{equation}
V(M)=-{1 \over 2}(T_c-T)M^2+{1 \over 4}M^4
\end{equation}
and is displayed in Fig.2d. Here, $V(M)$ is such that its maxima and minima correspond to unstable and stable equilibria, respectively. This potential function, and the dynamics it captures match several key properties of the underlying energy functions of the thermodynamic description\footnote{The original conceptualization is due to Lev Landau's approach to phase transitions, which simplifies the problem by focusing on an order parameter $\phi$, like magnetization or density, that summarizes the system's macroscopic state. The idea is that the system's free energy, which determines its stability, can be expressed as a smooth function of this order parameter.
Near the critical temperature $T_c$, the free energy is expanded as a series, with the terms chosen to satisfy the constraints imposed by the system's symmetry. In the magnetic transition, the term proportional to $M^2$ depends on $T-T_c$, so it changes sign at the critical temperature, while higher-order terms such as $M^4$ ensure the stability of the free energy for large $M$.}.

%%%%%%%%%%%%%%%%%%%%%%%%%%% BOX 1  %%%%%%%%%%%%%%%%%%%%%
\vspace{0.2truecm}
\noindent\fcolorbox{white}{MyGray}{
  \parbox{0.95\columnwidth}{
    \textbf{BOX 1. Symmetry-simplicity, broken symmetry-complexity.}

    \smallskip
\begin{small}
    Symmetry breaking is a fundamental mechanism that allows the emergence of complexity in natural systems \cite{anderson1972more,anderson1987broken,sole1996phase,smith2016origin,bar2019dynamics,krakauer2023symmetry}. Complexity arises when these symmetries are broken, allowing for differentiation, specialization, and the emergence of patterns in space and time.

    In the physical world, symmetry breaking explains how simple, homogeneous states give rise to diverse, ordered structures, such as convection cells emerging once a critical temperature is achieved \cite{velarde1980convection,nicolis1989exploring}. In developmental biology, broken symmetries were suggested by Rashevsky \cite{rashevsky1940physicomathematical} and Turing \cite{turing1952chemical} as the mechanism to generate large scale patterns out of microscopic interactions \cite{prigogine1967symmetry,nicolis1981symmetry}. In the early universe, symmetry breaking played a critical role during phase transitions, such as the separation of the fundamental forces \cite{mazumdar2019review}. These transitions allowed the initially uniform energy distribution to diversify, eventually leading to the formation of matter and the structured cosmos we observe today. Broken symmetries are also responsible for the generation of information \cite{nicolis1989exploring,haken2006information}.
\end{small}
  }
}
  \vspace{0.2truecm}

We can appreciate in a very intuitive way the presence and meaning of the broken symmetry: for $T<T_c$, any of the two possible attractors is identical, and which one is chosen is a path-dependent contingent event \cite{nicolis1989exploring}. The slightest deviation from $M^*=1/2$ (a random initial condition in the Ising model) will be amplified by the nonlinear character of the dynamics in an irreversible fashion. This important result has been discussed within the context of contingency in evolution \cite{gould1989wonderful,sole2024fundamental} and will have relevant consequences for our examples regarding the origins of life. 

The fundamental difference between a bifurcation and a thermodynamic phase transition lies in the nature of the systems they describe. A bifurcation occurs within a finite-dimensional state space, where the dynamical system possesses limited degrees of freedom. In contrast, a thermodynamic phase transition necessitates the thermodynamic limit, requiring the system size to approach infinity for singularities in thermodynamic quantities to emerge at critical points \cite{qian2016framework,bose2019bifurcation}. Similarly, the asymptotic limit $t \rightarrow \infty$ is essential for a dynamical system to reach its attractor state. Although this asymptotic time limit in bifurcations bears a conceptual resemblance to the thermodynamic limit in equilibrium phase transitions, the two remain mathematically and physically distinct \cite{goldenfeld2018lectures}.

%======================================================

\section{Chirality: a symmetry breaking scenario}

%======================================================

Many organic molecules exist in two forms that are mirror images of each other, somewhat like your left and right hands. Importantly, the mirror images cannot be superimposed: there is no way to twist the right hand so that it becomes a left one. The handedness of life is dramatically illustrated by looking at biochemical spaces and the representative molecules, such as DNA (Fig. 3a). For proteins, each aminoacid has two possible mirror images, usually indicated as $L-$ and $D-$ forms. With one exception (glycine), they are all chiral: the two images are not equivalent\footnote{Louis Pasteur discovered chirality in 1848 while studying tartaric acid crystals. He observed that solutions of this acid rotated polarized light, but synthetic tartaric acid did not. He examined the crystals under a microscope and noticed they existed in two mirror-image forms. He manually separated the two types and showed that each rotated light in opposite directions, revealing molecular asymmetry. This was the first evidence of molecular chirality} (Fig. 3b). Interestingly, living molecules are made of monomers with the same chirality: proteins are made of L-aminoacids, whereas nucleic acids are made of D-nucleotides: mirror symmetry is broken in living things \cite{hegstrom1990handedness}. Why is that?

The importance (and puzzle) of handedness has intrigued scholars from various disciplines \cite{saito2013colloquium}, including physics \cite{salam1991role,avetisov1991handedness,avetisov1996mirror}, chemistry\cite{sallembien2022possible}, molecular biology \cite{tverdislov2017periodic}, and, more recently, astrobiology \cite{chyba2005astrobiology}. Its relevance becomes clear when looking at polymers. Polymers were likely to be a crucial component of early life, allowing both catalysis and information storage, and they will play a key role in all our examples. In abstract terms, we build polymers of size $\nu$ from an alphabet $\Sigma$ of single units (monomers, such as amino acids or nucleotides). If the size of the alphabet s is indicated as $\vert \Sigma \vert$, the number of possible polymeric sequences scales as
\begin{equation}
{\cal N} \sim \vert \Sigma \vert^{\nu}
\end{equation}
Some simple examples illustrate the hyper-astronomic nature of these spaces. For aminoacids ($\vert \Sigma \vert=20$) forming peptides of length $\nu=100$, we obtain ${\cal N} = 20^{100}\sim 1.2 \times 10^{130}$. If we consider nucleotides (now with $\vert \Sigma \vert=4$) and a small sequence of size $\nu=10^2$, we have ${\cal N} = 4^{10^2}\sim 1.6 \times 10^{60}$. Considering a binary alphabet and assuming that both L and D possibilities are equally likely, the probability $P_L$ of having an all-L small chain of size $\nu=20$ is:
\begin{equation}
P_L = \left ( {1 \over 2 }\right )^{20} \sim 10^{-6}
\end{equation}
i.e. one in a million. This is a tiny chance. Polymers made of mixed $L$ and $D$ monomers could be argued to work, but early experiments revealed that polymers such as RNA grown from mixtures of monomers L and D fold in very different ways. Since functionality is based on the specific features of the folded structure, the required accuracy would be lost. Similarly, mixtures do not allow a key mechanism to copy information-carrying polymers: template-based replication. To guarantee that polymers are made of the same chiral monomers, they must already be present in the environment in just one form. To achieve such a handedness, some mechanism has to break the symmetry. 

Several mathematical models have been suggested to explain the broken symmetry in chiral molecules. The simplest one, initially formulated by Frank \cite{frank1953spontaneous,jafarpour2015noise,blackmond2019autocatalytic,laurent2021emergence,laurent2022robust,brandenburg2021homochirality} and several experimental studies analyzed, following these theoretical models \cite{soai1995asymmetric,sato2004asymmetric,islas2005mirror}. In its simplest form, we can formulate Frank's model using two types of chemicals, indicated by $D$ and $L$, which correspond to the two (chiral) forms. They can react with an additional molecule $A$ following the set of reactions: 
\begin{eqnarray}
A + D \stackrel{\mu}{\longrightarrow} 2D \; \; \; \; \; 
A + L \stackrel{\mu}{\longrightarrow} 2L  \; \; \; \; \; 
D + L \stackrel{\beta}{\longrightarrow} 2A
\end{eqnarray}
If we indicate the concentrations of the two forms by $D$ and $L$, we can derive the equations describing the dynamics of this mixture and study them using linear stability analysis.

\begin{figure*}[!t]
{\centering 
\includegraphics[width=\textwidth]{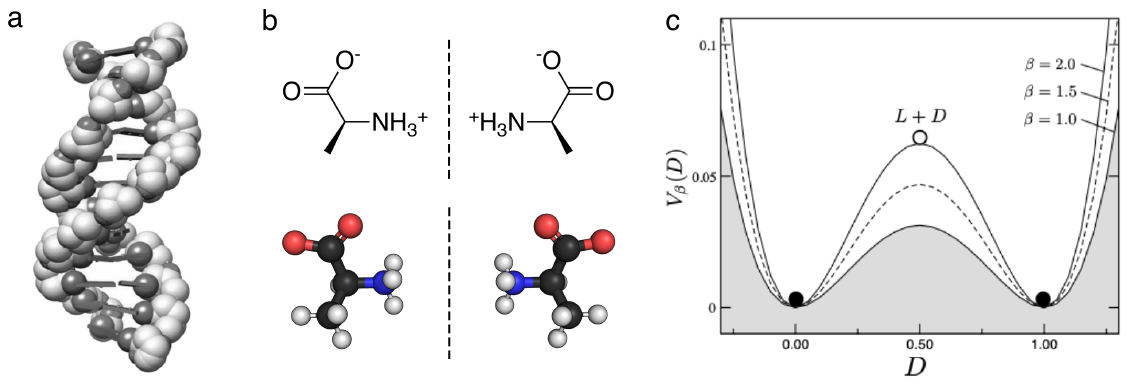}
\caption{Symmetry breaking and the origin of homochirality. Biological molecules have a well-defined chirality; for example, nucleic acids (DNA represented in 3D) always have the D form (while proteins are left-handed). This comes from the combination of many units having the same chirality. In (b), the aminoacid Alanine is shown in its two mirror configurations. Using Frank's model (see text), the potential function $V_{\beta}(x_1) = - \int f_{\beta}(x_1) dx_1 =\beta(x_1^4/2-x_1^3+x_1^2/2)$ associated to the dynamics of a racemic mixture is displayed (for different values of $\beta$). The unstable point (open circle) is associated with the $x_1=1/2=x_2$ racemic mixture, 
which is unstable. Deviations from this symmetric state lead to either $[D]=0,[L]=1$ or 
a $[D]=1,[L]=0$ final state.}
}
\end{figure*}

Let us indicate as $x_1=[D]$ and $x_2=[L]$ and consider the special case where the total population is conserved. Assuming this {\it constant population constraint} (CPC) i. e. $[D]+[L]=1$  we have:
\begin{eqnarray} 
{dx_1 \over dt}=  \mu x_1 - \beta x_1 x_2  - x_1 \Phi (x_1,x_2) \\
{dx_2 \over dt}=  \mu x_2 - \beta x_1 x_2  - x_2 \Phi (x_1,x_2) 
\end{eqnarray}
where the first two terms in the right-hand side correspond to the formation of 
molecules of each type and their conversion in $A$. The last terms introduce a dilution associated with an outflow. From the CPC condition, we have: 
$d (x_1+x_2)/dt=0=dx_1/dt+dx_2/dt$, which gives: $\Phi(x_1,x_2)=\mu-2\beta x_1 x_2$ and it can be shown that using $x_1=[D]$ as our macroscopic variable, we have a one-dimensional dynamical system:
\begin{equation}
{dx_1 \over dt}= f_{\beta}( x_1) = \beta x_1 (1-x_1)(2x_1-1)
\end{equation}
(a symmetric solution exists for $x_2=[L]$). The three fixed points are now: $x_1^* \in \{0,1,1/2\}$. 
The first two are stable, homochiral states, whereas the third corresponds to an unstable racemic. The stability condition from $\lambda(x_1)=\beta(6x_1-6x_1^2-1)$ gives now:
\begin{equation}
\lambda(0)= -\beta <0 \; \; \; \; \;\; \; \; \; \; 
\lambda \left ({1 \over 2} \right )= {\beta \over 2} >0  \; \; \; \; \;\; \; \; \; \;
\lambda(1)= -\beta <0
\end{equation}
we can see that the two chiral states are stable, whereas the racemic state is unstable. 
This means (as happened with the ferromagnetic dynamical model) 
a {\em symmetry breaking} takes place (refs)  where two alternative stable
states $x_1 = 0,1$ are possible, both accessible from $x_1=1/2$ through an amplification phenomenon. 
This can be seen using the so called potential function $V_{\beta}(x_1)$ defined again from:
\begin{equation}
{dx_1 \over dt}= - {d V_{\beta}(x_1) \over d x_1}
\end{equation}
The specific form of $V_{\beta}(x_1)$ is shown in Fig. 3c. Here the (unstable) racemic 
mixture $(D+L)$ and the two alternative (stable) homochiral configurations 
are displayed as empty and filled circles, respectively. Once we slightly deviate from the perfect racemic mixture, the ball rolls down towards one of the alternatives: the symmetry is broken towards a given chiral configuration. 

Is chirality a frozen accident? Is it a result of a contingent event, with no selection processes choosing one over the other potential outcomes? The previous model provides the simplest explanation for homochirality due to a path-dependent broken symmetry. However, some source of bias cannot be discarded \cite{deamer2011first}. The model, as defined, assumes an initial state that is fully symmetric (a perfect racemic mixture). Things would be different if this condition is not met, since even a small imbalance would be amplified. An interesting observation in this context comes from the analysis of meteorite aminoacids. These bodies carry organic matter \cite{oro1971amino,deamer1989amphiphilic} and it has been speculated that prebiotic molecular precursors have been seeded by comets and carbonaceous meteorites \cite{oro1961comets,chyba1990cometary,sephton2002organic}. Some evidence indicates that slight biases towards L-aminoacids are present in these meteorites \cite{engel1982distribution} but not in asteroids \cite{naraoka2023soluble,glavin2025abundant}. If present, the origin of these deviations is still under debate \cite{chieffo2023origin}.

%==========================================================================

\section{Replicators, quasispecies and the error threshold}\label{sec:eigen}

%==========================================================================

In this section, we consider the problem of understanding the population dynamics of early replicator molecules that undergo mutation and selection. Possible case studies would include populations of peptides or nucleic acids replicating and competing for resources in a prebiotic scenario. From now on, we will ignore the exact details of how such replication takes place, and we assume a polymer structure that can be described as a string of monomers of a given size $\nu$. Another relevant assumption can be made regarding mutation: in a prebiotic context, no error correction process is expected to operate (since no repair mechanisms were in place), and a fixed mutation rate $\mu_b$ per monomer and replication cycle is assumed. This is also the case for RNA viruses, which lack repair enzymes \cite{domingo2012viral}. Because mutation provides a source of molecular diversity, understanding the dynamics of these systems requires considering the sequence space (already mentioned above). A simple example is represented in Fig. 4a-c for a $\nu=4$ string size. If we indicate by ${\cal S}=(s_1,..,s_4)$ a given polymer made of monomers within the (binary) alphabet $\Sigma = \in \{ 0,1 \}$, the sequence space is a 4-dimensional hypercube ${\cal H}_4$. A given sequence will be located on a vertex of this cube, and the edges connecting two of these nodes indicate a one-mutation distance, and the size of this space will be $\vert {\cal H}^{\nu} \vert = \Sigma^{\nu}$. 

To define a selective advantage (allowing to define selection) we consider a sequence-fitness measure defined as:
\begin{equation}
f:\Sigma^{\nu}\longrightarrow U\subset{\cal R},
\end{equation}
which maps the string into a scalar value, \emph{i.e.}, 
\begin{equation}
{\cal S} \in \Sigma^{\nu} \longrightarrow f({\cal S})
\end{equation}
which will be defined in terms of replication rate of that given sequence. When a given sequence replicates, a new copy will be generated, provided no mutations occur, and a new mutant sequence will be created if it does not. Let us consider a special case described in \cite{swetina1982self}  where the sequence $(1,1,1,1)$, hereafter the {\it master sequence}, has a large fitness $f_(1,1,1,1)=f_m$ and all other sequences will have the same, but smaller fitness, namely $f({\cal S}\ne(1,1,1,1))=f\ll f_m$. This is often called a single-peak landscape. If we start from only master sequences, and represent the relative population size $x({\cal S}_i) \in \cal H$ of a given string ${\cal S}_i$ with a sphere with a radius proportional to population size.
As we increase mutation rates, more and more new strings will be mutants, thus reducing the size of the master sequence population while spreading towards neighboring nodes (fig.4b-c). Eigen dubbed this cloud of sequences a {\it quasispecies}. Intuition tells us that, since $f_m \gg f$, some master sequences would always be present. That is not the case: there is a phase transition at some critical mutation rate separating a phase where, indeed, the master sequence is present from another phase where populations are randomly spread across $\cal H$. In this context, the population size of the master sequence will be our order parameter whereas mutation will act as the control parameter. 

%In 1971, Manfer Eigen predicted a critical mutation rate, $\mu_c \sim 1/\nu$, beyond which Darwinian selection fails and random drift dominates for $\mu > \mu_c$. Below this threshold, information remains stable. Data confirms this inverse relationship, with mutation rates decreasing as genome length increases.

The loss of the master sequence and its consequences can be studied using replicator dynamics models incorporating mutation and selection. 
The Eigen-Schuster quasispecies model \cite{eigen1988molecular} considers a set of populations $\{ x_i \}$ representing the abundance of different genomes, changing in time by the following set of dynamical equations:
\begin{equation} 
{d x_i \over dt} = \sum_{j=1}^n f_j \, \mu(j \rightarrow i) \, x_j - \Phi({\bf x},t ) \, x_i ,
\end{equation}
where $x_i$ indicates the fraction of the population associated to the $i-$th mutant genome is very large, $\nu$ being the length of the genome). We assume CPC, with $\sum\limits_{j=1}^n x_j = 1$. Here $f_j$ is the growth rate of the $j$-th mutant, $\mu(j \rightarrow i)$ is the probability of having a mutation from sequence $j$ to sequence $i$ and $\Phi({\bf x})$\footnote{$\Phi({\bf x})$ is usually termed as the dilution outflow, which ensures a constant population ($\sum\limits_{i=1}^n x_i =1$  and $\sum\limits_{i=1}^n \dot{x}_i = 0$) also introducing competition between replicating genomes.} is the average fitness associated to the population vector ${\bf x}=(x_1, ... x_n)$, \emph{i.e.},
\begin{equation} 
\Phi({\bf x},t) = \sum_{j=1}^n f_j x_j \left ( \sum_{i=1}^n \mu(j \rightarrow i) \right ) = \sum_{j=1}^n f_j x_j =\langle f \rangle.
\end{equation}
For simplicity, let us consider the previous single-peak landscape described above (fig.4a-c) where the master sequence replicates more efficiently than any other sequence. Assuming that  \emph{i.e.}, $\mu(i \rightarrow j) = \mu$, we can split our system of equations into two sets: the master sequence and the mutant sequences. The system presented below is only a simplified approximation to the space connecting different genomes. A more accurate picture is provided by a given population $x_j$, mutation will be difficult to occur. In this case, we will have $x_m+x=1$ where we use $x=\sum_j x_j$. For the master sequence, we get: 
\begin{equation} 
{d x_m \over dt} = f_m(1-\mu)x_m + \sum_{j=1}^n f_j {\mu \over n} x_j - x_m \Phi({\bf x}),
\end{equation}
whereas, assuming that $f_{j \neq m}=f$, the set of equations for the mutant sequences reads: 
\begin{equation} 
{d x_i \over dt} = {f_m \mu \over n} x_m + 
f (1-\mu) x_i + \sum_{j \ne i}^n {f \mu \over n} x_j  - x_i \Phi({\bf x})
\end{equation}
It is easy to show, after some algebra, that the first equation can be simplified to
\begin{equation} 
{d x_m \over dt} \approx f_m(1-\mu)x_m - x_m \Phi({\bf x}),
\end{equation}
where we have used $\mu/n \ll 1$. Similarly, the aggregated population dynamics of mutant sequences is given by:
\begin{eqnarray} 
\sum_{i=1}^n {d x_i \over dt} ={f_m \mu \over n}\sum_{i=1}^n  x_m + f (1-\mu)\sum_{i=1}^n  x_i 
+\nonumber\nonumber\\
 {f \mu \over n} \sum_{i=1}^n \left ( \sum_{j \ne i}^n x_j \right ) - \sum_{i=1}^n x_i \Phi(t )
\end{eqnarray}
Given the homogeneous mutation rates, for large $n$ we obtain:
\begin{equation} 
{dx \over dt} \approx \mu f_m x_m + f x - x\Phi(t),
\end{equation}
the two previous differential equations for $x$ and $x_m$, along with the condition $x_m+x=1$ leads to an equation for the master sequence:
\begin{equation} 
{d x_m \over dt} = x_m \left [ (1-x_m)(f_m - f) - \mu f_m \right ].
\end{equation}

\begin{figure*}[!t]
\centering 
\includegraphics[width=\textwidth]{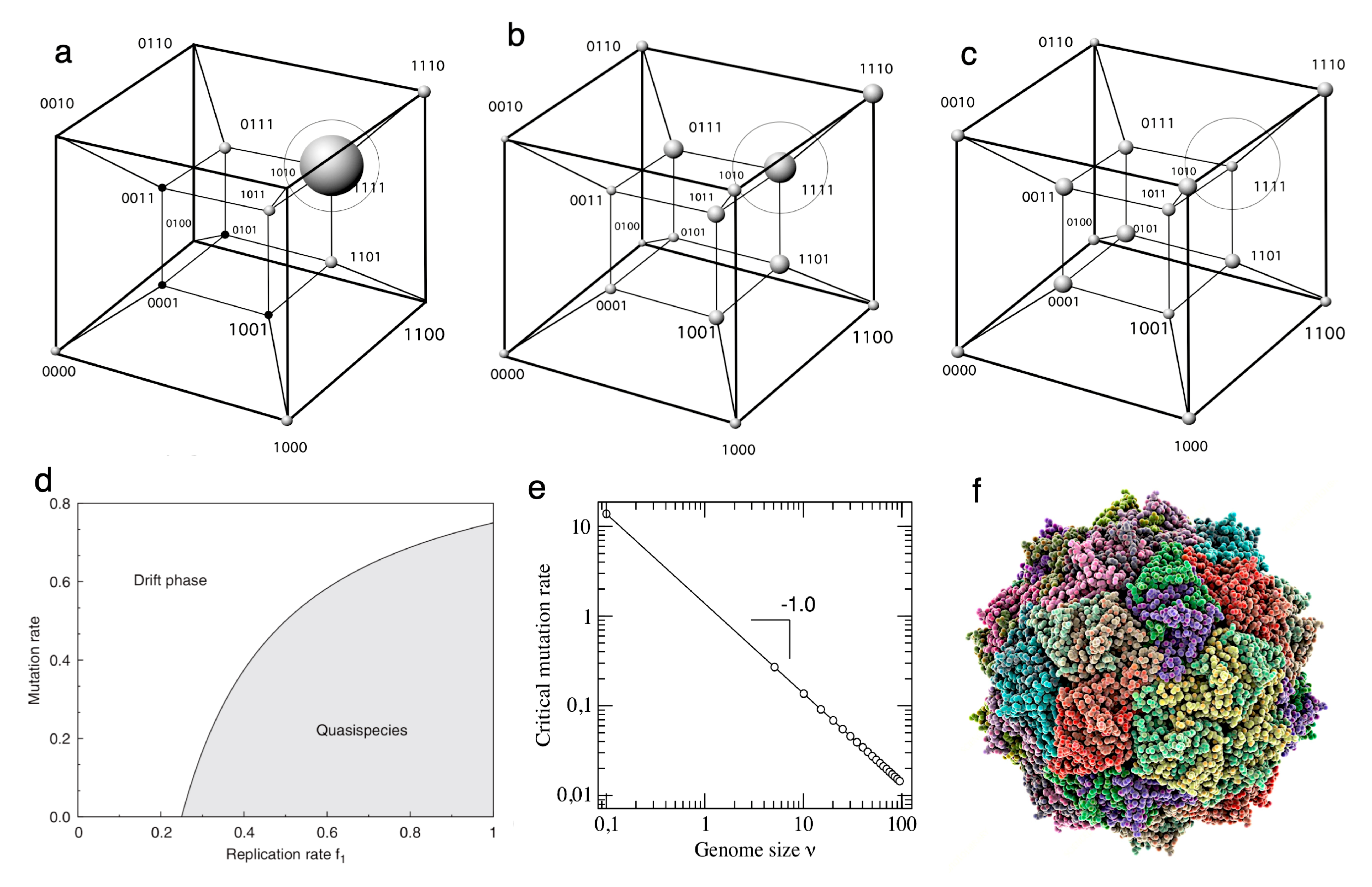}
\caption{The error threshold in self-replicating systems with mutation. Figures (a-c) display the fitness landscape for a $\nu=4$ genomes, ${\cal H}_4$, using different mutation rates: (a) sub-critical (the master sequence, located at $(1,1,1,1)$ has a large population with other strings around displaying marginal values), (b) close to critical (the master sequence is still present, with a wide cloud of mutants) and (c) supercritical with populations randomly wandering around (we have genetic drift). Two phases can be found in the replication-mutation parameter space (d): the quasispecies phase, where selection is at work, and the drift phase, where sequences wander around and a loss of information occurs. In (e) the predicted inverse scaling law between genome size and mutation rate is shown. In (f), the Q$\beta$ RNA virus is shown. This virus uses a replicase that is error-prone and has been used to explore the potential for simple self-replicating systems to function and evolve outside of cellular environments with minimal components.}
\end{figure*}

Two alternative equilibria are possible. The first one is $x_m=0$, which corresponds to the extinction of the master sequence and the full dominance of the pool of mutants, and a nontrivial equilibrium
\begin{equation}
x_m = 1 - \frac{\mu f_m}{f_m - f},
\end{equation}
allowing the coexistence of both master and mutant sequences. The later solution will be positive (and the master sequence will be present) provided that $x_m>0$, and this will occur if the mutation rate is lower than the critical value: 
\begin{equation} \label{error}
\mu < \mu_c = 1 - {f \over f_m}
\end{equation}

Fig. 4d shows the boundary separating the two phases. In the quasispecies phase, master genomes coexist with mutants. Crossing the boundary results in the disappearance of master sequences, leaving only mutants (grey area in Fig.4d). 
A very important scaling law connects the mutation rate of a genome to its length, which can be easily derived from our mean-field theory. Let us now consider the mutation rate per nucleotide.  it is related to $\mu$ by 
$\mu = 1-(1-\mu_b)^\nu$. Since $\mu_b$ is typically very small, we have: 
$\mu \approx 1-e^{- \mu_b \nu} \approx \mu_b \, \nu.$ If we return to the previous critical condition for mutation rates and write it down as a function of $\mu_b$ it turns out
\begin{equation}
\mu_b^c = {\alpha \over \nu},
\end{equation}
where $\alpha>0$ is a scaling factor. The last expression corresponds to the observed inverse decay of mutation rates as an inverse of their genome size, as shown in Fig.4e. This prediction was proven right once the estimation methods for mutation rates were available and obtained for RNA viruses. An example is the Q$\beta$ phage, shown in Fig.4f, and extensively used by Eigen et al. to study the experimental evolution of quasispecies \cite{eigen1988molecular}.

The universal character of the error threshold for our simple fitness landscape is highlighted by the mapping between a microscopic version of our previous model and the 2D Ising model. It turns out that the behaviour of both models are fundamentally equivalent \cite{eigen1988molecular,volkenstein2012physical}. The microscopic description of the quasispecies is obtained by finding the transition rates between digital genomes, given the structure of the sequence space $\cal H$ and the underlying fitness function. If $S_k$ and $S_j$ are two sequences of length $\nu$, the probability that a sequence $S_k$ replicates and mutates to a new sequence $S_j$ is given by:
\begin{equation}
W_{jk} = f(S_k)(\mu_b(1-\mu_b))^{\nu-d_H(S_j,S_k)}\mu_b^{d_H(S_j,S_k)}
\end{equation}
where the term $d_H(S_j,S_k)$ is the so-called Hamming distance between the two sequences (i.e. how far away they are within $\cal H$:
\begin{equation}
d_H(S_j,S_k)= {1 \over 2} \left [ \nu - \sum_{i=1}^{\nu} S_j^i S_k^i \right ]
\end{equation}
Using these expressions, we can simulate our quasispecies dynamics by explicitly considering the stochastic dynamics of mutation-selection. Importantly, the microscopic model correctly captures the existence of the error threshold and the phase transition for the single-peak landscape, as described above. A statistical physics approach reveals something remarkable. If we re-write $W_{jk}$ using the definition of $D_H$, the probability now reads
\begin{equation}
W_{jk} = f(S_k)[\mu_b(1-\mu_b)]^{\nu/2}\exp \left ( -K \sum_{i=1}^{\nu} S_j^i S_k^i \right )
\end{equation}
where $K=\ln(\mu_b/(1-\mu_b))/2$. Interestingly, we can notice the presence of a term (the sum on the right-hand side) that is already familiar to us: the energy function of the Ising model. It has been shown that the last expression is, in fact, identical to the so-called transfer matrix of the 2D Ising model,  which encodes the interactions between neighbouring rows of spins on the 2D lattice and is a key tool for solving the model \cite{leuthausser1986exact,leuthausser1987statistical,volkenstein2012physical}. 

Further developments within the modelling of life origins using statistical physics considered generalizations of these fitness functions using spin glass Hamiltonians, with energy functions including diverse couplings, i.e. a spin glass Hamiltonian ${\cal H} = -\sum_{i,j} J_{ij} S_iS_j$ \cite{stein1992spin,stein2013spin}. These generalized models allow one to explore more complex problems regarding the origins and evolution of early polymers \cite{anderson1983suggested,kauffman1993origins,peliti2002quasispecies}. Similarly, deterministic and stochastic models \cite{saakian2006quasispecies,saakian2006exact,sardanyes2010error} have been developed, as well as their spatially extended counterparts \cite{altmeyer2001error,pastor2001field,aguirre2008effects,sardanyes2008simple}. In this context, it is important to note that in rugged or complex fitness landscapes, where multiple peaks and valleys exist due to epistatic interactions between genes, the error threshold is not as straightforward, and the loss of information due to high mutation rates may not always result in complete meltdown. Instead, populations may shift between peaks or form diverse mutant distributions, and this has been explored under a generalized approach considering stochastic dynamics in finite population models \cite{nowak1989error} (see also \cite{demetrius1985polynucleotide,alves1998error,derrida1991evolution}). In such cases, the traditional notion of a sharp error threshold may blur into a more gradual transition, influenced by landscape ruggedness and selection strength.

Furthermore, neutral networks (i.e. clusters of genotypes with equivalent fitness connected by single mutations) despite high mutation rates, genetic information can persist without being strictly confined to a single sequence \cite{huynen1996smoothness}. This suggests that in highly complex landscapes, the threshold may depend on mutation rates and the structure of neutral spaces within the fitness landscape. In this context, experimental studies in viral evolution \cite{domingo2012viral,domingo2019viral} and computational models of evolutionary dynamics \cite{wilke2005quasispecies} further support the idea that error thresholds in complex landscapes manifest differently than in simple peak models. Viruses, for instance, often operate near their error threshold, but their adaptability is enhanced by mechanisms such as mutational robustness and recombination, allowing them to withstand mutation rates higher than predicted by simple models and mutation rates, with extinction thresholds different from the expected error threshold \cite{tejero2010effect,manrubia2010pathways}.  On the other hand, mutation rates can also be evolvable and respond to selection \cite{duffy2018rna,stich2007collective}.

%======================================================
%======================================================

\section{Autocatalysis, percolation and cycles}\label{sec:kauffman}

%======================================================
%======================================================

Our third class of models regarding phase transitions in early life scenarios concerns the emergence of biochemical cycles. Three classic papers describe different model approaches, all of them involving some kind of sef-organization process. These are Manfred Eigen's theory of hypercycles \cite{eigen1978hypercycle}, Stuart Kauffman, who introduced the conceptual framework of autocatalytic cycles \cite{kauffman1971cellular,kauffman1993origins,hordijk2019history} and Tibor Ganti's Chemoton model of a simple protocellular organization \cite{ganti1971principle,ganti2003chemoton,szathmary2005evolutionary}. Here, we explore the first two proposals and explain how their importance lies largely in the presence of complexity thresholds that can be described as a phase transition. Eigen’s hypercycle model addressed the problem of how self-replicating molecules could store and transmit genetic information despite high replication error rates. He proposed a network of mutually catalytic molecules, where each molecule enhances the replication of the next, creating a cooperative system that promotes stability and allows for the evolution of complexity. Kauffman's modelling approach, on the other hand, introduced the concept of autocatalytic sets, networks of molecules that collectively catalyze their formation. He argued that such networks could spontaneously arise in sufficiently diverse chemical environments, providing a plausible pathway for the origin of metabolism before the emergence of genetic systems \cite{hordijk2018autocatalytic}.

\subsection{Hypercycles: beyond the error threshold}

The error threshold defined by Eigen's theory of quasispecies imposes an upper bound to the complexity of early molecular replicators. Lacking error-correcting mechanisms, there seems to be no way out to keep expanding informational complexity. Moreover, this theoretical framework is defined within the context of replicator equations \cite{hofbauer1998evolutionary,nowak2006evolutionary} that assume competitive interactions among different kinds of molecules. 
Consequently, we have two important negative effects influencing the diversity of molecular networks, namely (a) the presence of a sharp limit to genome length and (b) the reduction of diversity due to competitive exclusion. These two limitations affect two crucial requirements to evolve complexity: individual complexity and ecological diversity.

One solution to this problem (which effectively presents a paradox) was provided by cooperation, a common factor in the emergence of evolutionary novelty \cite{schuster1996does}. Specifically, it was suggested that such an innovation could be achieved by the {\it hypercycle}, i.e. a catalytic network consisting of self-replicating units, each capable of catalyzing the replication of another unit in the network, forming a closed loop \cite{eigen1977principle,eigen1978hypercycle,eigen1978hypercycleC}. An example is shown in Fig 5a, where a $n$-member hypercycle defines a closed loop of catalytic reactions. The set of nonlinear equations can describe this reads:
 \begin{equation}
{d x_i \over dt} =  \Gamma_{i,i-1} x_i x_{i-1} - x_i \sum_{j=1}^n \Gamma_{j,j-1} x_j x_{j-1}
\end{equation}
where the $i$-th member of the cycle catalyses the growth of the $(i+1)$-th species while it is also 
helped by the $i-1$-th one in the loop (with the constraint $i+1 \rightarrow n$ when $i=n$) \cite{eigen1977principle}. As we can see, the hypercycle incorporates a major evolutionary change: the presence of cooperative interactions. Here, molecular interactions enhance the chances to replicate other molecules, which has been linked to the early RNA molecules acting as both enzymes and templates \cite{higgs2015rna,papastavrou2024rna} (see also \cite{copley2007origin}). The simplest example of this class of cooperative loop is the two-member hypercycle (fig. 5b), which has been extensively studied . We can connect both kinds of dynamical systems, namely competitive and cooperative dynamics, using a two-dimensional model where two populations exhibit both competitive and cooperative interactions. This can be represented as a dynamical system as follows: 
 \begin{eqnarray}
{d x_1\over dt} &=&  r_1 x_1 + \Gamma x_1 x_2 - x_1 \Phi(x_1,x_2)\;\;\;\;\;\;\;\;\;\\
{d x_2\over dt} &=&  r_2 x_2 + \Gamma x_1 x_2 - x_2 \Phi(x_1,x_2)
 \end{eqnarray}
where the two molecular species, with relative abundances $x_1$ and $x_2$, grow and interact in two ways, once again assuming CPC (i.e. $x_1+x_2=1$). While each population can grow with some intrinsic rate $r_i$ (hereafter, we consider $r_1>r_2$), there is also a cross-interaction term $\Gamma x_i x_j$ indicating the presence of some cross-catalysis. Why this model? If we ignore the cooperative term (i.e. $\Gamma=0$), we have a so-called replicator model dominated by competition. Using the CPC condition, it is easy to show that a logistic model is obtained, namely the dynamics for $x_1$:
 \begin{equation}
{d x_1 \over dt} =  (r_1-r_2)x_1(1-x_1)
\end{equation}
and it is easy to see that the only nontrivial fixed point is $x_1^*=1$, and thus $x_2^*=0$: competitive exclusion (and thus diversity reduction) occurs. Conversely, if we take $r_i=0$, a purely cooperative system is obtained. In this case, the resulting 1D model reads: 
 \begin{equation}
{d x_1 \over dt} =  \Gamma x_1(1-x_1)(1-2x_1)
\end{equation}
which has the same three fixed points as our example of chirality above. However, in this case, the exclusion points $ x_1*=0,1$ are unstable, whereas the coexistence one, i.e. $ x_1*=1/2$, is now stable. 
How do we move from the competitive scenario, where only one genome will persist, to the rich, cooperative catalytic loop that allows the persistence of both (and thus a much larger information capacity)? Let us take the full model with both terms and analyse its bifurcation diagram:
 \begin{equation}
{d x_1 \over dt} =  2\Gamma x_1(1-x_1) \left ({r_1-r_2+\Gamma \over 2\Gamma }- x_1 \right )
\end{equation}
and the corresponding potential function:
\begin{eqnarray}
V(x_1) &=& - \int {d x_1 \over dt} dx_1 \nonumber\\
&=&  -\Gamma {x^4_1 \over 2}- \left( \Gamma + {r_1-r_2 \over 3} \right)x_1^3-\nonumber\\
&&(r_1-r_2+\Gamma){x_1^2 \over 2}
\end{eqnarray}
In this case, the two trivial solutions $x_1^*=0$ and $x_1=1$
are unstable and stable, respectively, provided that $r_1-r_2=\Gamma_c<\Gamma$, whereas the coexistence solution will be stable for $\Gamma>r_1-r_2$. This is illustrated in Fig.5c, where we show the bifurcation diagram displaying $x_1^*$ against $\Gamma$ and the corresponding potentials are outlined in Fig. 5d for different parameter values. We use $r_2=0.2$ and $r_1=0.3$. Below $\Gamma_c$, competition dominates and the fastest replicator wins ($x_1^*=1$ is stable). However, once we cross this critical value, a sharp change is observed: the stable point is no longer a homogeneous population, but a mixture of the two. This illustrates the point that we can have a punctuated evolutionary event once a threshold is crossed, allowing for a new complexity level when new dynamical rules dominate the dynamics. Importantly, the hypercycle has also been shown to overcompete Malthusian replicators: once the hypercycle is established, cannot be easily replaced by a newcomer \cite{eigen1978hypercycle,michod1983population}. Eigen's hypercycle has inspired a whole generation of models of prebiotic evolution, particularly in connection with complex dynamics \cite{schnabl1991full,stadler1992mutation,sardanyes2006ghosts,guillamon2015bifurcations}, spatial patterns \cite{boerlijst1991spiral,szabo2002silico,sardanyes2006bifurcations,sardanyes2007spatio} or information \cite{eigen1981origin,kuppers2012molecular,schuster2016increase}  and provided the theoretical framework for experimental paths towards molecular hypercycles \cite{sole2016synthetic,higgs2015rna,adamski2020self,mizuuchi2022evolutionary}. Finally, because of its special growth properties, it has been conjectured that a hypercyclic form of evolutionary dynamics might have been a requirement for the propagation of life in the biosphere \cite{wilkinson2007fundamental,conde2020synthetic}. 

This cyclically connected system has also been proposed as a critical step in transitioning from non-living to living chemistry\cite{eigen1977principle} where more complex networks of reactions and biochemical cycles occur \cite{xavier2020autocatalytic}. To illustrate this concept and its connection with a phase transition, let us consider, once again, a toy model of population dynamics.

%%%%%%%%%%%%%%%%%%%%%%%%%%% BOX 2  %%%%%%%%%%%%%%%%%%%%%%%%%
\vspace{0.2truecm}
\noindent\fcolorbox{white}{MyGray}{
  \parbox{0.95\columnwidth}{
    \textbf{BOX 2. Emergence of system-level structures: percolation}

    \smallskip
\begin{small}
Percolation in random graphs describes how connections between nodes in a network form and grow, leading to a sudden transition from small, isolated clusters to a large, interconnected system when a critical density of connections, known as the percolation threshold, is reached. In the context of the origins of life, this concept illustrates how random interactions between molecules in a "primordial soup" could have initially created small, isolated networks, which eventually merged into a giant, interconnected system capable of supporting self-sustaining chemical processes like replication and metabolism. This sudden emergence of large-scale connectivity provides an intuitive framework for understanding how complexity and order can arise spontaneously from random interactions, with recent work showing why the sparsity of connected networks is another emergent feature of real-world systems~\cite{ghavasieh2024diversity}. Mathematically, the existence of this phase transition can be demonstrated by using a random graph $\Omega=(V,E)$ made of a set $V$ of $N$ nodes (vertices) and a set of edges $E$ connecting them. If we indicate as $k_i$ the number of links of a given node, such network's average connectivity (degree) will be $\langle k \rangle$. If $p$ indicates the probability that two nodes are connected, 
we have $\langle k \rangle=pN$. It can be shown \cite{newman2006random,sole2011phase} that a critical connectivity $k_c=1$ (or $p_c=1/N$ separates two well-defined phases. For $p<p_c$, a disconnected set of small networks, whereas for $p>p_c$, we find a system where most elements are connected to others, forming a so-called giant component. 

\end{small}
  }
}
\vspace{0.2truecm}

\begin{figure*}[t]
    \centering
    \includegraphics[width=18 cm]{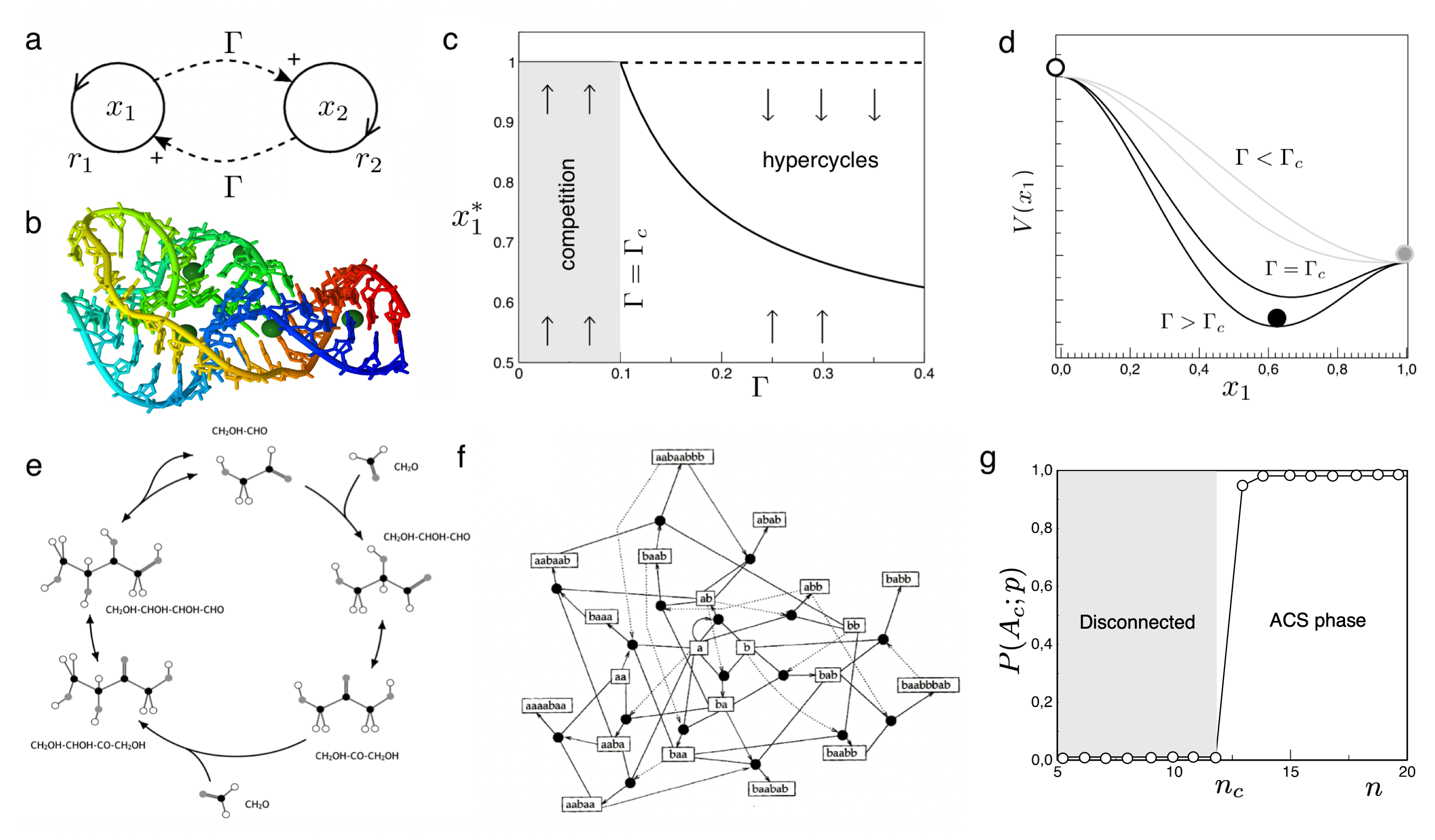}
    \caption{{\textbf Phase transitions in molecular cycles.} The two-member hypercycle (a) is a simple model of cooperative interactions where two molecular species replicate by helping each other (using template replication, for example). Some molecular candidates are small RNA molecules such as the hammerhead ribozyme (b) that can act as both a template and an enzyme. A simple model, where each species can replicate at some rate $r_i$ and cooperate at a rate $\Gamma$, reveals a phase transition (c) separating a domain of competition-dominated dynamics from a phase where cooperation allows the coexistence of both species. Here, we use $r_1>r_2$, and the critical point is given by $\Gamma_c=r_1-r_2$. The associated potential for this model is displayed in (d) for subcritical (grey lines) and supercritical phases (solid lines). Catalytic cycles have been found in a few organic chemistry contexts, such as the Formose reaction (e), and complex cycles can emerge within a rich network of molecules. In (f), an example is shown for a model of polymers of different lengths and a binary alphabet. A phase transition occurs in systems involving many molecules with a probability $p$ acting as catalysis of a given reaction. In (g), an example is shown, where an algorithm was built to detect the presence and characteristics of ACS in random catalytic networks \cite{hordijk2004detecting}. Two transition curves are shown, indicating the probability of finding a catalytic cycle.
    }
    \label{fig:ACSexamples}
\end{figure*}

\subsection{Autocatalytic cycles}

The previous example illustrates the role critical thresholds play in promoting molecular complexity due to catalytic interactions among two different molecules. Higher complexity (and information) was achieved in the hypercycle cooperation. Cycles are common within biological systems and might be a universal feature of life \cite{morowitz1966physical,sole2024fundamental}. This is particularly obvious for metabolism, where cycles allow for the efficient recycling of intermediates, enabling cells to conserve energy and resources. They provide a continuous flow of reactions that generate essential molecules, such as ATP, while maintaining balance in metabolic pathways. They also integrate different inputs and outputs, ensuring metabolic flexibility and adaptability to changing environmental or cellular conditions. Some examples of catalytic cycles have been described within organic chemistry, such as the so-called Formose reaction \cite{niitsu1992analysis,cleaves2023formose}. This reaction (Fig. 5e) is of interest in prebiotic chemistry as a possible pathway for the synthesis of simple sugars. Catalytic cycles have also been engineered using template replication using nucleotide-based oligomers \cite{sievers1994self} and RNA-based networks \cite{vaidya2012spontaneous}. However, they can also emerge from inorganic chemistry, as shown in \cite{miras2020spontaneous}, where cooperative cycles generate complex nanostructures based on molybdenum. Some early studies in this context have suggested that such cycles could have emerged through crystallization in a mineral context \cite{cairns1966origin}.

We can think of metabolic cycles as one case study within the potential reaction networks that can be built from a set of molecular species. A general dynamical model for a population of interacting molecules that catalyze each other's synthesis has been used traditionally within the context of artificial chemistries \cite{dittrich2001artificial,banzhaf2015artificial} and chemical reaction dynamics \cite{stadler1993random}. One standard formulation reads: 
\begin{equation}
    {dx_k \over dt}  = \sum_{i =1}^s \sum_{j=1}^s \alpha^k_{ij}x_j x_i -x_k \Phi({\bf N})
\label{eq:qs}
\end{equation}
where $x_i$ is the population of type $i$, ${\bf x}=(x_1, ..., x_s)$, and each sum includes the different potential reactions allowing the production of type $k$, i.e., the bimolecular reactions $i + j \longrightarrow i+j+k$ 
 with their corresponding rates $\{ \alpha^k_{ij}\}$. These are kinetic models grounded in reactions assuming random molecule collisions. Once again, we have a term 
%$\Phi({\bf N},t)$ 
$\Phi({\bf x})$ that keeps the condition 
%$\sum_{j \in\mathcal{S}} N_j =1$
$\sum_{j} x_j =1$.
In our modern biosphere, metabolic pathways are run using enzymes that speed up chemical reactions, making them occur fast enough to sustain life \footnote{They lower the activation energy needed for reactions, allowing metabolic processes to be efficiently regulated and controlled, ensuring cells can produce energy and vital molecules when needed.}. However, how can cycles emerge in a prebiotic soup where enzymes were not yet in place and catalysis might have been inefficient?

Stuart Kauffman (motivated by the emergence of aminoacids in Miller experiments) investigated the problem of how self-sustaining cycles could occur in a prebiotic scenario \cite{kauffman1986autocatalytic,kauffman1993origins,farmer1986autocatalytic,dyson1982model,bagley1990spontaneous,bagley1992evolution,filisetti2012stochastic}. Kauffman's concept of autocatalytic sets is a foundational idea in understanding the origin of life and self-organization in complex systems \cite{bagley1989modeling,hordijk2019history}. The idea revolves around the notion that life could emerge from a network of molecules that catalyze the formation of each other in a mutually reinforcing manner. These autocatalytic sets are collections of molecules where each member is produced through reactions catalyzed by other members of the set. Importantly, the set as a whole becomes self-sustaining under the right conditions. In chemical reaction networks, molecules are represented as nodes and chemical reactions form the links between them (Fig. 5f). Kauffman proposed that, as the number of molecules and reactions increases, there is a critical point—akin to a phase transition—at which large, connected networks of reactions are inevitable, including autocatalytic sets. Kauffman’s model uses random graph theory to describe this phenomenon, as described in the classical work by Paul Erd\"os and collaborators \cite{erdos1960evolution,bollobas1998random}. In a random reaction network, the likelihood that a reaction exists between two molecules is determined by a probability $p$. As 
$p$ increases, the network transitions from having small, disconnected clusters of reactions to forming a single, large, connected component. This critical transition (Box 2) corresponds to the point where the system could achieve sufficient connectivity for autocatalytic cycles to emerge.

In its original formulation, the model considers a population of strings (polymers) such as simple autocatalytic protein networks, where amino acid sequences catalyze the joining (ligation) of shorter sequences and the breaking (cleavage) of longer ones. Let us label as $\Omega$ the set of possible polymers, where each chain $\omega_k \in \Omega$ is described by a string $\omega_k=\omega_k^1,...,\omega_k^m$, where $m =2,3,...,n$ and $\omega_k^j \in \Sigma$, where for simplicity we assume $\Sigma=\{ 0,1 \}$. As defined, the set of strings is defined by the space 
\begin{equation}
    {\cal S} = \Sigma^2 \cup \Sigma^3 \cup ...\cup \Sigma^n
\end{equation}
What is the phase transition point connecting the $p$ parameter with the molecular diversity of our system? Here we summarize the derivations from \cite{hordijk2004detecting,mossel2005random,hordijk2019molecular}. The number of molecular types is given by 
\begin{equation}
    \vert   {\cal S} \vert = \sum_{k=2}^n \Sigma^k = 2^{n+1}-2
\end{equation}
and we can now use the critical condition from random graphs, which established that the system will display a very large connected fraction of reactions on a single, large graph, namely that $|E|/|V|>1/2$, we can derive the corresponding expression for our system. If $\cal R$ is the set of reactions, it can be shown that the inequality becomes: 
\begin{equation}
    p \vert   {\cal R} \vert = p(n-2)2^{n+1} > {1 \over 2}
\end{equation}
the critical point for a fixed $n$ is thus: 
\begin{equation}
    p_c = {1 \over (n-2)2^{n+1}} 
\end{equation}
or, alternatively, if we fix $p$, the critical $n$ will be obtained by solving the equation:
\begin{equation}
    n_c + \log_2 n_c = \log_2 \left ( {1 \over p} \right ) - 2
\end{equation}
An example of the phase transition for a system of strings of different maximal length $n$, using $p \sim 10^{-5}$ is shown in Fig. 5g (for details see \cite{hordijk2019molecular}). For this particular $p$, the estimated $n_c$ from theory is $n_c \sim 12$, consistent with the simulation results.

Conceptually, autocatalytic sets illustrate how order and function can emerge spontaneously in systems governed by simple rules, challenging the idea that life required highly improbable events to originate. Kauffman's work emphasizes the importance of networks and interactions over individual components, providing insights into the nature of complexity and the possible pathways to the first living systems.

%\section{Prevolutionary transitions}\label{sec:nowak}

%%%%%%%%%%%%%%%%%%%%
%%%%%%%%%%%%%%%%%%%%
%%%%%%%%%%%%%%%%%%%%
\section{Discussion}\label{sec:conclusions}

How did life originate? Was life a rare event, unlikely to be seen in other planets? The transition from living to non-living matter is one of the greatest scientific challenges, and its solution will require the use of interdisciplinary works that combine multiple disciplines. From Miller's experiments to the search for biosignatures in exoplanets, research on the origins of life has been expanding its horizon. Over the last decades, an emergent field, astrobiology, has been building around the goal of a synthesis of knowledge across disciplines and scales. Better experimental tests, powerful instruments searching for biosignatures and statistical tests to validate them have been developed. Along with these efforts, dedicated studies have been pushing the theoretical side of this enterprise. Among the key problems being addressed, we need to understand how crucial events in the path to living matter took place. This goal could benefit from theoretical frameworks where some universal traits might be identified beyond the specific components at work. Theoretical models based on phase transitions and nonlinear dynamics offer complementary insights, explaining how complexity emerges through processes like symmetry breaking, cooperation, and network formation. Together, experimental and theoretical approaches provide a unified framework to understand life’s emergence and its potential universality.

In this paper, we have chosen three examples that illustrate this approximation: the handedness of biological molecules, the presence of limits to the molecular complexity of error-prone replicators, and the emergence of complex networks of cooperative molecular assemblies. These three case studies have been modeled using population dynamics equations and the observed regularities explained by means of well-known phenomena, including symmetry breaking, percolation and cooperative exchanges. Broken symmetries have been known to pervade complexity in many different contexts, and might have played a key role in shaping biological asymmetries in our early biosphere. 

%%%%%%%%%%%%%%%%%%%%%%%%%%% BOX 3 %%%%%%%%%%%%%%%%%%%%%%%%%
\vspace{0.2truecm}
\noindent\fcolorbox{white}{MyGray}{
  \parbox{0.95\columnwidth}{
    \textbf{BOX 3. Thermodynamics of evolution and the origin of life}

    \smallskip
\begin{small}
Interestingly, Vanchurin et al~\cite{vanchurin2022thermodynamics} proposed to describe the origin of life as a phase transition in which a system of interacting molecules shifts to an ensemble of self-replicating organisms, driven by thermodynamic and statistical principles. The control parameter governing this transition is the evolutionary temperature $T$, which quantifies the level of stochasticity in the evolutionary process: for decreasing $T$ selective forces become more effective, stabilizing complex and self-sustaining structures. In this context, the order parameter could be plausibly interpreted as the fraction of trainable, adaptive heritable variables $q_a$ that have become effectively stabilized as core variables $q_c$, representing the shift from a dynamic ensemble of molecules to structured, self-replicating entities.
In the prebiotic phase, molecular interactions were governed by transient chemical affinities, with adaptation driven by environmental fluctuations rather than encoded memory. As the system crosses the transition threshold, a growing fraction of variables undergo stabilization, leading to the emergence of persistent genetic information. The adaptive variables include genetic sequences, regulatory interactions, and metabolic components, which become subject to selection and drive long-term evolution. Indicating with $\Omega_{p}(\mathcal{T},\mathcal{M})$ the physical grand potential -- where $\mathcal{T}$ and $\mathcal{M}$ indicate the physical temperature and chemical potential -- and with $\Omega_{b}(T,\mu)$ the biological grand potential -- where $T$ and $\mu$ indicate the evolutionary temperature and evolutionary potential -- the transition is formally described by $\Omega_{p}(\mathcal{T},\mathcal{M})=\Omega_{b}(T,\mu)$, marking a shift in thermodynamic constraints. At the corresponding critical point, the free energy landscapes of the molecular and organismal phases become equivalent, where the system undergoes a qualitative reorganization in its statistical and thermodynamic properties. This transition also reflects a deeper principle, where learning dynamics at multiple levels generate hierarchical structures, enabling the evolution of complexity through successive phase transitions~\cite{vanchurin2022toward}.
\end{small}
  }
}
\vspace{0.2truecm}

Populations of replicators unable to correct errors due to mutation, on the other hand, will jeopardise their complexity by errors, and limited by the presence of an error threshold that imposes a limit to sequence length. The transition found here has deep connections with canonical models of physics, such as the Ising model, hinting to deeply universal connections between them. Finally, the emergence of complex molecular networks, including those able to overcome the error catastrophe (the hypercycles) is inevitable, as described by the presence of a percolation threshold separating a molecular soup of disconnected components from a connected one where catalytic cycles can emerge. In all these examples, the macroscopic patterns described provide a rationale to the origins of complexity that paved the way to life.

By framing the origin of life within the context of phase transitions and emergent behaviours, this work establishes a robust theoretical foundation for understanding abiogenesis and early life evolution. This perspective not only unifies disparate phenomena under a cohesive framework but also offers a pathway for guiding future experimental investigations. The fundamental mechanisms that govern transitions in nonlinear dynamical systems provide critical insights into the universal laws underlying the emergence of life. These principles hold the potential to determine whether life is an inevitable outcome of complex systems or a rare, highly localized phenomenon.

\vspace{6pt}

\appendix

\acknowledgments
RS thanks Stuart Kauffman, Kepa Ruiz Mirazo, Steen Rasmussen, Andreea Munteanu, Susanna Manrubia, Josep Sardanyes, Ben Shirt-Ediss, Jordi Piñero, Norman Packard, John McCaskill, and Peter Schuster for conversations about life origins over the years. RS was supported by an AGAUR FI-SDUR 2020 grant and by the Santa Fe Institute. MDD acknowledges partial financial support from the Human Frontier Science Program Organization (HFSP Ref. RGY0064/2022).

% Bibliography
\bibliographystyle{apsrev4-1}
\bibliography{biblio}  % Assumes you have a 'biblio.bib' file

\end{document}